\documentstyle[aps,epsf]{revtex}
 
\newcommand{\bq}{\begin{equation}}
\newcommand{\eq}{\end{equation}}
\newcommand{\bqn}{\begin{eqnarray}}
\newcommand{\eqn}{\end{eqnarray}}
\newcommand{\nb}{\nonumber}
\newcommand{\lb}{\label}

\begin{document} 
\title{Levi-Civita Solutions Coupled with Electromagnetic Fields}
\author{A.Y. Miguelote \thanks{E-mail: yasuda@dft.if.uerj.br}, 
 M.F.A. da Silva \thanks{E-mail: mfas@dft.if.uerj.br},
Anzhong Wang \thanks{E-mail: wang@dft.if.uerj.br}}
\address{  Departamento de F\' {\i}sica Te\' orica,
Instituto de F\' {\i}sica, Universidade do Estado do Rio de Janeiro,
 Rua S\~ ao Francisco Xavier 524, Maracan\~ a,
20550-013 Rio de Janeiro~--~RJ, Brazil}
\author{N.O. Santos \thanks{E-mail:
nos@lacesm.UFSM.br and nos@lafex.cbpf.br}} 
\address{Laboratorio de Astrofisica e Radioastronomia,
Centro Regional Sul de Pesquisas Espaciais - INPE/MCT,
Cidade Universitaria, 97105-900, Santa Maria, RS, Brazil
\\
and \\
LAFEX, Centro Brasileiro de pesquisas F\'{\i}sicas - CBPF, Rua Dr. Xavier
Sigaud 150, Urca, CEP: 22290-180, Rio de Janeiro~--~RJ, Brazil}

\date{\today}

\maketitle

\begin{abstract}

 The local and global properties of the Levi-Civita (LC) solutions  coupled with 
an electromagnetic field are studied and some limits to the vacuum LC
solutions are given. By doing such limits, the  physical and geometrical
interpretations of the free parameters involved in the solutions are made
clear. Sources for both the LC vacuum solutions and the LC solutions coupled
with an electromagnetic field are studied, and in particular it is found that 
all the LC vacuum solutions with $\sigma \ge 0$ can be produced by
cylindrically symmetric thin shells that satisfy all the energy conditions,
weak, dominant, and strong. When the electromagnetic field is present, the
situation changes dramatically. In the case of a purely magnetic field, all
the solutions with $\sigma \ge 1/\sqrt{8}$ or  $\sigma \le - 1/\sqrt{8}$ can
be produced by physically acceptable cylindrical thin shells, while in the
case of a purely electric field, no such shells are found for any value of
$\sigma$.

\end{abstract}   
 
\section{Introduction}

Spacetimes with cylindrical symmetry have been studied intensively in the
past twenty years or so, in the context of topological cosmic strings that may
have formed in the  early stages of the Universe \cite{VS94}, and in
gravitational collapse \cite{AT92,PW00}. Recently, the physical and geometrical
interpretations of the Levi-Civita (LC) vacuum solutions, which represent the
most general cylindrical static vacuum spacetimes, has attracted much
attention \cite{Bonnor92}. In particular,  it has been shown, among
other things, that the LC vacuum solutions can be produced by cylindrically
symmetric   sources, which satisfy all the energy conditions, weak, strong,
and dominant \cite{HE73}, only for $0 \le \sigma  \le 1$ \cite{WSS97}, where
$\sigma$ is one of the two parameters appearing in the LC vacuum  solutions,
which is related to, but in general not equal to the mass per unity length 
\cite{Bonnor92}. It has been also shown that when the LC solutions coupled
with a cosmological constant, the spacetime structures are dramatically
changed, and in some cases they give rise to  black hole structures
\cite{SWPS00}. 

In this paper, our purpose is two fold. First, as we mentioned above, so far,
physically acceptable and cylindrically symmetric sources for the LC vacuum
solutions are found only  for $0 \le \sigma  \le 1$. Since $\sigma = 1$ does
not represent any typical value in these solutions \cite{WSS97}, it has
troubled us for a long time why the solutions with $\sigma > 1$ cannot be
realized by physically acceptable cylindrical sources. In this paper, we shall
show that all the LC vacuum solutions can be produced  by cylindrically
symmetric thin shells that satisfy all the energy conditions, as long as
$\sigma \ge 0$. The key observation that leads to such a conclusion is that,
in the both limits, $\sigma   \rightarrow 0, \;  + \infty$, the solutions
become (locally) Minkowskian but with different identifications of the angular
and axial coordinates. Thus, it is very plausible that, as $\sigma $
increases to a certain value, the axial and angular coordinates may change
their roles. By constructing cylindrically symmetric sources, in this paper we
shall confirm this claim, and argue that the change should happen at $\sigma =
1/2$, although in the range $1/4 \le \sigma \le 1$, physically acceptable
sources for both of the two identifications are found. It is remarkable to
note that it is exactly this range that timelike circular geodesics do not
exist \cite{GH69,Luis01}. Thus, the interpretation of all the LC vacuum
solutions with $\sigma \ge 0$ as representing cylindrically symmetric vacuum
spacetimes is physically acceptable. It is interesting to note that several
authors already speculated that as $\sigma$ increases to the value $\sigma =
1/2$, the angular coordinate should be straightened out to infinite, so that the
resultant spacetimes  become plane symmetric \cite{GH69}. Indeed,  we have
shown that the solution with $\sigma = 1/2$ can be produced by a massive plane
with an uniform distribution of matter \cite{SWS98}, while Philbin has shown
that all the solutions with $|\sigma| > 1/2$ can be produced by  massive planes
\cite{Ph96}.  However, as far as we know, this is the first time to be argued
that, when $\sigma > 1/2$, the two spacelike coordinates $x^{2}$ and $x^{3}$
change their roles, and show that the LC vacuum solutions can be produced by
cylindrical sources for $\sigma \ge 0$, after such an exchange of coordinates
is taken place. The physics that provokes such an exchange is not understood,
yet. Second, we shall extend   our studies to the LC solutions coupled
with electromagnetic fields and study the effects of the electromagnetic
fields on the local and global structure of the spacetimes. 

The paper is organized as follows: In Sec. II, we
shall study the main properties of the LC solutions when coupled with an
electromagnetic field, and take their vacuum limits. By doing so, we can find
out the   physical and geometrical interpretations of the free parameters
involved in the solutions. In Sec. III, we shall consider cylindrically
symmetric thin shells that produce the spacetimes described by the LC vacuum
solutions  or by the LC solutions coupled with an electromagnetic field. We
use Israel's method \cite{Israel66} to obtain the general expression for the
surface energy-momentum tensor of a thin shell, which separates two arbitrary
cylindrical static regions. Then we  apply these general formulae to the case
where the shell separates a Minkowski-like internal region from an external
region described by either the LC vacuum solutions or the LC solutions coupled
with an electromagnetic field. Imposing the energy conditions, we  show that
only for some particular choices of the free parameters appearing in the
solutions these conditions are  fulfilled. The paper is closed with Sec. IV,
where our main conclusions are presented.

\section{ Levi-Civita Solutions Coupled with Electromagnetic Fields}

The static spacetimes with cylindrical symmetry are described by the metric
\cite{Kramer80}
\bq
\label{2.1}
ds^2 = f(R){dT}^ 2 - g(R){dR}^{2} -
h(R)\left(dx^{2}\right)^{2} -
l(R)\left(dx^{3}\right)^{2},
\eq
where $T$ and $R$ are, respectively, the timelike and radial coordinate. In
general, the spacetime possesses three Killing vectors, $\xi^{\mu}_{(0)} =
\delta^{\mu}_{0}, \;  \xi^{\mu}_{(2)} = \delta^{\mu}_{2}$, and
$\xi^{\mu}_{(3)} = \delta^{\mu}_{3}$, where $\{x^{\mu}\} = \{T, \; R, \;
x^{2},\; x^{3}\}$. Clearly, the coordinate
transformations 
\bq
\lb{2.1a}
T = a \tilde{T},\;\;\; 
R = R (\tilde{R}),\;\;\;
x^{2} = \alpha \tilde{x}^{2},\;\;\;
x^{3} = C^{-1} \tilde{x}^{3}, 
\eq
preserve the form of metric,  where $a, \; \alpha$ and $C$ are arbitrary
constants, and $R (\tilde{R})$ is an  arbitrary function of the new radial
coordinate $\tilde{R}$. A spacetime with cylindrical symmetry must obey
several   conditions \cite{Kramer80,PSW96,MS98}:

(i) {\em The existence of an axially symmetric axis}: The spacetime that has
an axially symmetric axis is assured by the condition,
\bq
\lb{2.2}
||\partial_{\varphi}|| = |g_{\varphi\varphi}| \rightarrow O(R^{2}),
\eq
as $R \rightarrow 0^{+}$, where we had chosen the radial coordinate
such that the axis is located at $R = 0$, and $\varphi$ denotes the angular
coordinate with the hypersurfaces $\varphi = 0$ and $\varphi = 2\pi$ being
identical. Since both $\xi^{\mu}_{(2)}$ and $\xi^{\mu}_{(3)}$ are spacelike
Killing vectors, $\varphi$ can be chosen to be either $x^{2}$ or $x^{3}$. This
ambiguity always rises, since the Einstein field equations are differential 
equations, and consequently do not determine the global topology of the
spacetime. This observation will be crucial in understanding the LC
vacuum solutions to be discussed in Secs. II and III below.   However, once
$\varphi$ is identified, the rescaling transformation of Eq.(\ref{2.1a}) for
$\varphi$,   
\bq 
\lb{2.2a}
 \varphi = \frac{\tilde{\varphi}}{D}, 
\eq 
maps the two identified hypersurfaces $\varphi = 0$ and $\varphi =
2\pi$, respectively, to $\tilde{\varphi} = 0$ and $\tilde{\varphi} = 2\pi D$.
Consequently,  it results in an angular defect in the coordinates 
$\{T, \; R,\; \tilde{\varphi}, \; z\}$, given by
 \bq 
\lb{2.2b}
\Delta {\tilde{\varphi}} = 2\pi(1 - D).
\eq 
Thus, the coordinate transformation (\ref{2.2a}) in general 
yields physically different solutions.  
In particular, when the spacetime outside the axis is locally Minkowskian, this 
angular defect can be associated with a cosmic string located on 
the axis \cite{VS94}.

(ii) {\em The elementary flatness on the axis}: This condition requires that
the spacetime be locally flat on the axis, which in the present case can be
expressed as
\bq
\lb{2.3}
\frac{X_{,\alpha}X_{,\beta}g^{\alpha\beta}}{4X} =
\frac{g^{2}_{\varphi\varphi,R}}{4g_{RR}g_{\varphi\varphi}} \rightarrow 1,
\eq
as $R \rightarrow 0^{+}$, where $X$ is given by $X = ||\partial_{\varphi}|| =
|g_{\varphi\varphi}|$, and $(\; )_{,R} \equiv \partial
(\; )/\partial R$. Note that solutions that fail to satisfy this
condition are sometimes accepted since the appearance of spacetime
singularities on the axis can be considered as representing the existence of
some kind of sources \cite{Bonnor92}. For example, when the left-hand side of
Eq.(\ref{2.3}) approaches a finite constant, the singularity on the axis can
be related to a cosmic string \cite{VS94}.  

(iii) {\em No closed timelike curves}: In the cylindrical spacetimes, closed
timelike curves (CTCs) are rather easily introduced \cite{HE73}. While the
physics of the CTCs is not yet clear \cite{Hawking92}, we shall not consider
this possibility in this paper and simply require that
\bq
\lb{2.4}
g_{\varphi\varphi} < 0,
\eq
hold in all the region of the spacetime considered. 

(iv) {\em Asymptotical flatness}: When the sources are confined within a
finite region in the radial direction, one usually also requires that the
spacetime be asymptotically flat as ${\cal{R}} \rightarrow + \infty$,  where
${\cal{R}}$ denotes the geometric proper distance from the axis to a referred
point in the radial direction. 

It should be noted that because of the
cylindrical symmetry, the spacetime can never be asymptotically flat in the
axial direction. Therefore, in the following whenever we mention that the
spacetime is asymptotically flat, it always means that it is  asymptotically
flat only in the radial direction.

For an electromagnetic field $A_{\mu}(R)$, the energy-momentum tensor (EMT) is
given by
\bq
\lb{2.5}
T_{\mu\nu} = \frac{2}{\kappa}\left(
F_{\mu\alpha}F_{\beta\nu}g^{\alpha\beta}  + \frac{1}{4}g_{\mu\nu}
F_{\alpha\beta}F^{\alpha\beta}\right),
\eq
where $\kappa (\equiv 8\pi G/c^{4})$ is the Einstein gravitational coupling
constant, and
\bq
\lb{2.6}
F_{\mu\nu} \equiv A_{\mu,\nu} - A_{\nu,\mu}.
\eq
Because of the symmetry, from Eq.(\ref{2.6}) we find that 
\bq
\lb{2.7}
F_{02} = F_{03} = F_{23} = 0.
\eq
On the other hand, when the electromagnetic field is source-free, we have
\bq
\lb{2.8}
F^{\mu\nu}_{\;\;\; ;\mu} = F^{1\nu}_{\;\;\; ,R} +
\frac{1}{2}\left[\ln(fghl)\right]_{,R} F^{1\nu}=0,
\eq
where the semicolon ``;" denotes the covariant derivative.  Clearly, the above
equation has the general solution 
\bq
\lb{2.9}
F^{1\nu} = \frac{B^{\nu}} {(fghl)^{1/2}},\; (\nu = 0, 1, 2, 3),
\eq
where $B^{\nu}$ are the integration constants with $B^{1} = 0$. 
Substituting Eqs.(\ref{2.7}) and
(\ref{2.9}) into Eq.(\ref{2.5}) we find that
\bqn
\lb{2.10}
T_{\mu\nu} &=& \frac{1}{\kappa fhl}\left\{
f\left[f\left(B^{0}\right)^{2} + h\left(B^{2}\right)^{2} 
+ l\left(B^{3}\right)^{2}\right]\delta^{0}_{\mu}\delta^{0}_{\nu}
- g\left[f\left(B^{0}\right)^{2} - h\left(B^{2}\right)^{2} 
- l\left(B^{3}\right)^{2}\right]\delta^{1}_{\mu}\delta^{1}_{\nu}\right.\nb\\
& &  + h\left[f\left(B^{0}\right)^{2} + h\left(B^{2}\right)^{2} 
- l\left(B^{3}\right)^{2}\right]\delta^{2}_{\mu}\delta^{2}_{\nu}
+ l\left[f\left(B^{0}\right)^{2} - h\left(B^{2}\right)^{2} 
+ l\left(B^{3}\right)^{2}\right]\delta^{3}_{\mu}\delta^{3}_{\nu}\nb\\
& & \left. - 2 B^{0}B^{2}fh\left(\delta^{0}_{\mu}\delta^{2}_{\nu} +
\delta^{2}_{\mu}\delta^{0}_{\nu}\right) 
- 2 B^{0}B^{3}fl\left(\delta^{0}_{\mu}\delta^{3}_{\nu} +
\delta^{3}_{\mu}\delta^{0}_{\nu}\right) 
+ 2 B^{2}B^{3}hl\left(\delta^{2}_{\mu}\delta^{3}_{\nu} +
\delta^{3}_{\mu}\delta^{2}_{\nu}\right) \right\}.
\eqn
When the electromagnetic field is the only source for the Einstein field
equations, $G_{\mu\nu} = \kappa T_{\mu\nu}$, we find that the components
$T_{02},\; T_{03}$ and $T_{23}$ have to vanish, because the Einstein tensor
$G_{\mu\nu}$ for the metric (\ref{2.1}) has no non-diagonal terms. The vanishing
of these terms yields,
\bq
\lb{2.11}
B^{0}B^{2} = 0,\;\;\; B^{0}B^{3} = 0,\;\;\; B^{2}B^{3} = 0,
\eq
which have  four different solutions,  
 \bqn
\lb{2.12}
A) \; B^{0} &=&   B^{2} = B^{3} = 0,\;\;\;
B) \; B^{2} \not= 0, \;\;\; B^{0} = B^{3} = 0, \nb\\ 
C) \; B^{3} &\not=& 0, \;\;\; B^{0} = B^{2} = 0, \;\;\;\;
D) \; B^{0} \not= 0, \;\;\; B^{2} = B^{3} = 0.    
\eqn
In the following, let us consider them separately.

\subsection{ $B^{0} =   B^{2} = B^{3} = 0$}

In this case, the electromagnetic field vanishes and the corresponding
solution is the LC vacuum solution, given by \cite{Kramer80}
\bq
\lb{2.13}
ds^{2}  =  a^{2}R^{4\sigma}dT^{2} - R^{4\sigma(2\sigma
-1)}\left[dR^{2} + \alpha^{2}\left(dx^{2}\right)^{2}\right]
  - C^{-2}R^{2(1 - 2\sigma)}\left(dx^{3}\right)^{2},
\eq
where $a, \; \sigma, \; \alpha$ and $C$ are the integration constants. Without
loss of generality, one can always make $a = 1$ by rescaling
the timelike coordinate, $\tilde{T} = a T$, and assume that $\alpha$ and $C$
are all positive.  The physical meaning of $\alpha$ and $C$ depend on the
choice of the angular coordinate $\varphi$. For example, if $\varphi$ is
chosen as $x^{3}$, then $C$ will be related to the angular defect parameter
$D$, and $\alpha$ has no physical meaning and can be transformed away by the
rescaling $\tilde{x}^{2} = \alpha x^{2}$. However, if $\varphi$ is chosen as
$x^{2}$, then the roles of $\alpha$ and $C$ will be exchanged.   The parameter
$\sigma$ is related, but not equal, to the mass per unit length
\cite{Bonnor92}, and physically acceptable sources have been found so far only
for $ 0 \le \sigma \le 1$ \cite{WSS97}. When $\sigma = 0,\;  1/2$, the
corresponding solutions are flat in the region $0 < R < + \infty$. It was
shown that in the case $\sigma = 1/2$, the  $(x^{2}, x^{3})$-plane can be
extended to infinity, $ - \infty < x^{2}, \; x^{3} < + \infty$. Then, the 
resultant  spacetime has plane symmetry and can be produced by a massive plane
with uniform energy density \cite{SWS98}.   Thus, in the vacuum case there are
only two physically essential parameters, one is related to the mass per unit
length, and the other is related to the angular defects. 

Making the   coordinate transformations,    
\bq 
\lb{2.13a} 
\tilde{R} = \cases{(2\sigma -1)^{-2(2\sigma -1)^{2}/{\cal{A}}}  R^{(2\sigma
-1)^{2}},& $ \sigma \not= 1/2$,\cr 
\ln R, & $ \sigma = 1/2$,\cr}
\eq
 we find that the metric (\ref{2.13}) can be written as, 
\bq
\lb{2.13b}
ds^{2}  = \cases{\tilde{R}^{4\sigma/(2\sigma -
1)^{2}}\left(d\tilde{T}^{2} - d\tilde{R}^{2}\right) -  
\tilde{\alpha}^{2}\tilde{R}^{4\sigma/(2\sigma - 1)} \left(dx^{2}\right)^{2}  -
 \tilde{C}^{-2}\tilde{R}^{2/(1 - 2\sigma)}\left(dx^{3}\right)^{2},&  $\sigma
\not= 1/2$,\cr 
e^{2\tilde{R}} \left(d{T}^{2} - d\tilde{R}^{2}\right) -  
{\alpha}^{2} \left(dx^{2}\right)^{2}  -  {C}^{-2} \left(dx^{3}\right)^{2},& 
$\sigma = 1/2$,\cr} 
\eq
where ${\cal{A}} \equiv 4\sigma^{2} - 2\sigma + 1$, and
\bq
\lb{2.13c}
\tilde{T} \equiv (2\sigma - 1)^{4\sigma/{\cal{A}}}T,\;\;\;
\tilde{\alpha} \equiv \alpha (2\sigma - 1)^{4\sigma(2\sigma - 1)/{\cal{A}}},\;\;\;
\tilde{C} \equiv C (2\sigma - 1)^{2(2\sigma - 1)/{\cal{A}}}.
\eq
It is interesting to note that, in the limit $\sigma \rightarrow 0$, the
metric becomes locally Minkowskian with $x^{3}$ as the angular
coordinate and $x^{2}$ the axial coordinate, while as $\sigma \rightarrow +
\infty$, the metric  becomes also locally Minkowskian but now with $x^{2}$ as the
angular coordinate and $x^{3}$ the axial coordinate. This suggests
that there may exist a critical value $\sigma_{c}$, when $\sigma <
\sigma_{c}$, $x^{3}$ should be taken as the angular coordinate,   and when 
$\sigma > \sigma_{c}$, $x^{2}$ should be taken as the angular coordinate. The
analysis given below will confirm this speculation.

\subsection{$B^{2} \not= 0, \;\; B^{0} = B^{3} = 0$}

In this case, since the components $F_{0\mu}$ vanish, the corresponding
electromagnetic field is purely magnetic and produced by a  current
along the axis $x^{2}$ \cite{Bonnor54}. The corresponding EMT is given by  
\bq
\lb{2.14}
T_{\mu\nu} = \frac{\left(B^{2}\right)^{2}}{\kappa fl}\left(
f \delta^{0}_{\mu}\delta^{0}_{\nu}
+ g  \delta^{1}_{\mu}\delta^{1}_{\nu} 
+ h \delta^{2}_{\mu}\delta^{2}_{\nu} 
- l\delta^{3}_{\mu}\delta^{3}_{\nu}\right). 
\eq
Solving the corresponding Einstein field equations, we find that the solutions
are given by
\bqn
\lb{2.15}
f &=& g = R^{2m^{2}}G^{2},\;\;\;
h = \frac{\alpha^{2}}{G^{2}},\;\;\; 
l = \frac{R^{2}G^{2}}{C^{2}},\nb\\
F^{\mu\nu} &=& \frac{C B^{2}}{\alpha R^{2m^{2}+1}G^{2}}
\left(\delta^{\mu}_{1}\delta^{\nu}_{2} -
\delta^{\mu}_{2}\delta^{\nu}_{1}\right),\;\;\;  B^{2} = \pm
\left(\frac{4c_{1}c_{2}m^{2}}{C^{2}}\right)^{1/2}, 
\eqn 
where $G$ is given by
\bq
\lb{2.16}
G \equiv c_{1}R^{m} + c_{2}R^{-m},
\eq
and $\alpha, \; C, \; c_{1}, \; c_{2}$ and $m$ are integration constants.
Since $F_{\mu\nu}$ is real, we must have $c_{1}c_{2} \ge 0$ in the present
case. These solutions are actually  Witten's case 1 solutions
\cite{Witten62}. 

When $m = 0$, the electromagnetic field vanishes, and the
corresponding spacetime is locally Minkowskian,
\bq
\lb{2.17}
ds^{2} = G^{2} \left[dT^{2} - dR^{2}
- \frac{\alpha^{2}}{G^{4}} \left(dx^{2}\right)^{2}
- \frac{R^{2}}{C^{2}} \left(dx^{3}\right)^{2}\right],\; (m = 0),
\eq
where $G = c_{1} + c_{2}$. Clearly, now the angular coordinate $\varphi$
should be chosen to be $x^{3}$, and the axial coordinate $z$ to be $x^{2}$.
Then, the constant $C$ is related to the angular defect of the spacetime  
\cite{VS94}, and the constant $\alpha^{2}/G^{4}$ can be made disappear  by
rescaling $x^{2}$, while the conformal factor $G^{2}$ can be transformed away
by conformal transformations. Thus, in the above metric the only physically
essential parameter is $C$.

When  $c_{1} = 0, \; c_{2} \not=0$, we find that the electromagnetic field
vanishes identically,  $F^{\mu\nu} =0$, and the metric becomes,
\bq
\lb{2.18}
ds^{2} = c^{2}_{2} \left[R^{2m(m-1)}(dT^{2} - dR^{2})
- \frac{\alpha^{2}R^{2m}}{c^{4}_{2}} \left(dx^{2}\right)^{2}
- \frac{R^{2(1-m)}}{C^{2}} \left(dx^{3}\right)^{2}\right], \;(c_{1} = 0).
\eq
Thus, without loss of generality we can set $c_{2} = 1$. Then,
comparing Eq.(\ref{2.18}) with Eq.(\ref{2.13b}), we find that
\bq
\lb{2.19}
m = \frac{2\sigma}{2\sigma - 1},\;\; (c_{1} = 0,\; \sigma \not= 1/2).
\eq
It is interesting to note that in this case there is no direct limit of the
LC solution with $\sigma = 1/2$. 

When  $c_{1} \not= 0, \; c_{2} =0$,   we have
$F^{\mu\nu} =0$, too, and the corresponding metric and the constant $m$ can be
obtained from Eqs.(\ref{2.18}) and (\ref{2.19}) by replacing $c_{2}$ by
$c_{1}$ and changing the sign of $m$. 

When  $ c_{1}c_{2} \not= 0$, defining 
\bq
\lb{2.19ab}
c_{1} = \beta^{m}\gamma, \;\;\;\;
c_{2} = \frac{\gamma}{\beta^{m}},
\eq
where $\beta > 0$, we find that the corresponding metric takes the form
\bq
\lb{2.19b}
ds^{2} = \bar{\gamma}^{2}\bar{R}^{2m^{2}}G^{2}_{+}\left(d\bar{T}^{2} -
d\bar{R}^{2}\right) -
\frac{\bar{\alpha}^{2}}{G^{2}_{+}}\left(dx^{2}\right)^{2} -
\frac{R^{2}G^{2}_{+}}{{\bar{C}}^{2}}\left(dx^{3}\right)^{2},\;
(c_{1}c_{2} \not= 0), 
\eq
where
\bqn
\lb{2.19c}
\bar{R} &=& \beta R,\;\;\; \bar{T} = \beta T,\;\;\;
{G_{+}} \equiv \bar{R}^{m} +
\bar{R}^{-m},\nb\\
\bar{\alpha} &\equiv& \frac{\alpha}{\gamma},\;\;\; 
\bar{\gamma} \equiv \frac{\gamma}{\beta^{m^{2} + 1}},\;\;\;
\bar{C} \equiv \frac{\beta C}{\gamma}.
\eqn
Then, it can be shown that in the coordinates $\{\bar{x}^{\mu}\} = \{\bar{T},\;
\bar{R},\; x^{2},\; x^{3}\}$, the  electromagnetic field takes the form,
\bq
\lb{aaa} 
\bar{F}^{\mu\nu} = \pm \frac{2m}{\bar{\alpha} \bar{\gamma}^{2}
\bar{R}^{2m^{2}+1}G^{2}_{+}} \left(\delta^{\mu}_{1}\delta^{\nu}_{2} -
\delta^{\mu}_{2}\delta^{\nu}_{1}\right).
\eq
 
From the above analysis we can see that {\em when the LC solutions coupled with
an electromagnetic field, there are only three physically independent
free parameters, $m, \; \bar{\gamma}$, and one of the two parameters
$\bar{\alpha}$ and $\bar{C}$}, the latter depends on the choice of the
angular coordinate $\varphi$ \cite{Witten62}.  In the following studies of
these solutions we shall work only in the $\bar{x}^{\mu}$ coordinates, and  
drop all the bars from the constants and coordinates defined above, without
causing any confusions.   

It interesting to note that the spacetime remains
the same if we change the sign of the parameter $m$. Therefore, without loss
of generality,  we shall  consider only the case where $m \ge
0$.  To study the singularity behavior of the solutions, we find that  
\bqn 
\lb{2.20} 
F&\equiv&
F^{\alpha\beta}F_{\alpha\beta} =
\frac{8m^{2}}{\gamma^{2}R^{2(m^{2}+1)}G_{+}^{4}},\nb\\ 
I &\equiv& R^{\alpha\beta\gamma\sigma} R_{\alpha\beta\gamma\sigma}  =
\frac{16m^{2}}{\gamma^{4}R^{4(m^{2}+1)}G_{+}^{8}}\times\nb\\ & &
\left\{(m+1)^{2}\left[m(m+1) + 1\right] R^{4m} + (m-1)^{2}\left[m(m-1)
+ 1\right] R^{-4m}\right.\nb\\ & & 
\left. - 6m(m+1)^{2} R^{2m} + 6m(m-1)^{2} R^{-2m} -
2(m^{4} - 12m^{2} + 1)\right\},
\eqn 
from which we have
\bq
\lb{2.21}
F = \cases{\infty, & $m \not=  1$\cr
{8\gamma^{-2}}, & $ m =  1$ \cr}\;\;,\;\;\;\;\;
I = \cases{\infty, & $m \not=  1$\cr
-{320\gamma^{-4}}, & $ m =  1$ \cr}\;\;,
\eq
as $R \rightarrow 0^{+}$, and $F$ and $I$ all go to zero as $R
\rightarrow + \infty$. One can show that all the fourteen scalars built from
the Riemann tensor have similar behavior. Therefore, all these solutions
are asymptotically flat as $R \rightarrow +\infty$ and singular at $  R = 0$,
except for the one with $m =  1$. The singularities at $R = 0$ are timelike
and naked. The corresponding Penrose diagram for the solutions with $m \not=
 1$ is given by Fig. 1. 

When $m =  1$,   the metric (\ref{2.19b}) takes
the form 
\bq
\lb{2.22}
ds^{2} = \gamma^{2}\left(1 + R^{2}\right)^{2}
\left(dT^{2} - dR^{2}\right) -  \frac{\alpha^{2}R^{2}}{\left(1 +
R^{2}\right)^{2}}\left(dx^{2}\right)^{2} 
- \frac{\left(1 + R^{2}\right)^{2}}{C^{2}}\left(dx^{3}\right)^{2},\; (m = 1),
\eq
and the corresponding electromagnetic field is given by,
\bq
\lb{2.22aaa} 
{F}^{\mu\nu} = \pm \frac{2}{{\alpha} {\gamma}^{2}
{R} \left(1 + R^{2}\right)^{2}} \left(\delta^{\mu}_{1}\delta^{\nu}_{2} -
\delta^{\mu}_{2}\delta^{\nu}_{1}\right).
\eq

Clearly, in this case $x^{2}$ should be chosen as the angular coordinate. If
we further set $\alpha = \gamma$, then  the solution is locally flat on the
axis $R = 0$, asymptotically flat as $ R \rightarrow + \infty$, and free of
any kind of spacetime singularities in the whole spacetime. Thus, in this case
the spacetime is geodesically complete and the  corresponding Penrose diagram
is  given by Fig. 1, too, but now the vertical line $R = 0$ is free of   
spacetime singularities. Its applications to Cosmology were first studied by
Melvin \cite{Melvin64} and Thorne \cite{Thorne65}, and the model is usually
referred as to the Melvin Universe. This model has been extensively studied in
various theories recently, such as, $N=2$ Supergravity, heterotic string
theory, and Non-linear Electrodynamics \cite{GH01}.  

\subsection{$B^{3} \not= 0, \;\; B^{0} = B^{2} = 0$}

In this case, the only non-vanishing component of $F^{\mu\nu}$ is 
$F^{13}$, and the corresponding electromagnetic field is purely magnetic and
produced by a  current along the axis $x^{3}$.  The EMT given by
Eq.(\ref{2.10}) now becomes     
\bq
\lb{2.14a}
T_{\mu\nu} = \frac{\left(B^{3}\right)^{2}}{\kappa fh}\left(
f \delta^{0}_{\mu}\delta^{0}_{\nu}
+ g  \delta^{1}_{\mu}\delta^{1}_{\nu} 
- h \delta^{2}_{\mu}\delta^{2}_{\nu} 
+ l\delta^{3}_{\mu}\delta^{3}_{\nu}\right). 
\eq
Solving the corresponding Einstein field equations, we find that the solutions
are given by
\bqn
\lb{2.15a}
f &=& g = R^{2m^{2}}G^{2},\;\;\;
h = \frac{R^{2}G^{2}}{C^{2}},\;\;\; 
l = \frac{\alpha^{2}}{G^{2}},\nb\\
F^{\mu\nu} &=& \frac{C B^{3}}{\alpha R^{2m^{2}+1}G^{2}}
\left(\delta^{\mu}_{1}\delta^{\nu}_{3} -
\delta^{\mu}_{3}\delta^{\nu}_{1}\right),\;\;\;  B^{3} = \pm
\left(\frac{4c_{1}c_{2}m^{2}}{C^{2}}\right)^{1/2}, 
\eqn 
where $G$ is still given by Eq.(\ref{2.16}). These solutions are
actually Witten's Case 2 solutions \cite{Witten62}. Comparing
Eq.(\ref{2.15}) with Eq.(\ref{2.15a}), we find that if we exchange the two
coordinates $x^{2}$ and $x^{3}$, we shall get one solution from the other.
Hence, the physical and geometrical properties of these solutions can be
obtained from the ones given by Eq.(\ref{2.15})  by exchanging the
two coordinates $x^{2}$ and $x^{3}$.

\subsection{$B^{0} \not= 0, \;\; B^{2} = B^{3} = 0$}

In this case, the only non-vanishing component of $F^{\mu\nu}$ is 
$F^{01}$, and the corresponding electromagnetic field is purely electric and
produced by an  axial charge distribution. The spacetime in this case has
been studied by several authors  in different systems of
coordinates \cite{Muk38}. In the system of coordinates adopted in this
paper, the EMT given by Eq.(\ref{2.10}) now becomes     
\bq
\lb{2.14b}
T_{\mu\nu} = \frac{\left(B^{0}\right)^{2}}{\kappa hl}\left(
f \delta^{0}_{\mu}\delta^{0}_{\nu}
- g  \delta^{1}_{\mu}\delta^{1}_{\nu} 
+ h \delta^{2}_{\mu}\delta^{2}_{\nu} 
+ l\delta^{3}_{\mu}\delta^{3}_{\nu}\right), 
\eq
and the corresponding solutions are given by
\bqn
\lb{2.15b}
f &=& G^{-2}, \;\;\; l = \frac{R^{2}G^{2}}{C^{2}},\;\;\;
g = \alpha^{-2}h = R^{2m^{2}}G^{2}, \nb\\
F^{\mu\nu} &=& \frac{C B^{0}}{\alpha R^{2m^{2}+1}G^{2}}
\left(\delta^{\mu}_{0}\delta^{\nu}_{1} -
\delta^{\mu}_{1}\delta^{\nu}_{0}\right),\;\;\;  B^{0} = \pm
\left(\frac{- 4\alpha^{2}c_{1}c_{2}m^{2}}{C^{2}}\right)^{1/2}. 
\eqn  
Clearly, to have $F^{\mu\nu}$  real, now we must require $c_{1}c_{2} \le 0$.
The above solutions are Witten's Case 3 solutions \cite{Witten62}. This class
of solutions have been studied by several authors  using different
forms of metric \cite{Muk38}.  The  form used above is the same as in
\cite{Kramer80}. 

When $m = 0$, similar to Case B, the electromagnetic field vanishes, and the
corresponding spacetime is also locally Minkowskian,
\bq
\lb{2.17a}
ds^{2} = G^{2} \left[G^{-4}dT^{2} - dR^{2}
- \alpha^{2}  \left(dx^{2}\right)^{2}
- \frac{R^{2}}{C^{2}} \left(dx^{3}\right)^{2}\right],\; (m = 0),
\eq
where $G = c_{1} + c_{2}$. Clearly, now the angular coordinate $\varphi$
should be chosen to be $x^{3}$, and the axial coordinate $z$ to be $x^{2}$.
The  only physical constant is $C$ that is related to the
angular defect of the spacetime \cite{VS94}.    

When $c_{1} = 0,\; c_{2} \not= 0$, the
corresponding electromagnetic field vanishes, and the solutions reduce to 
\bq
\lb{2.23c}
ds^{2} =
c^{2}_{2}\left\{R^{2m}\left(\frac{dT}{c^{2}_{2}}\right)^{2} -
R^{2m(m -1)}\left[dR^{2} +  \alpha^{2} \left(dx^{2}\right)^{2}\right] -
C^{-2}R^{2(1-m)} \left(dx^{3}\right)^{2}\right\},\; (c_{1} = 0). 
\eq
Comparing it with Eq.(\ref{2.13}) we find that
\bq
\lb{2.25a}
m = 2\sigma,\; ( c_{1} = 0).
\eq

When $c_{1} \not= 0,\; c_{2} = 0$,  the
corresponding solutions can be obtained from Eq.(\ref{2.23c})  by replacing
$c_{2}$ by $c_{1}$ and by changing the signs of $m$. 

When $c_{1}c_{2} \not= 0$,   the solutions have
also three physically independent parameters. In fact, introducing the two
parameters $\beta$ and $\gamma$ via the relations, 
 \bq
\lb{2.19aa}
c_{1} = \beta^{m}\gamma, \;\;\;\;
c_{2} = -\frac{\gamma}{\beta^{m}},\;\; (\beta > 0),
\eq
we find that the corresponding metric takes the form
\bq
\lb{2.19bb}
ds^{2} =
\bar{\gamma}^{2}\left(\frac{d\bar{T}^{2}}{G_{-}^{2}}
- \bar{R}^{2m^{2}}G_{-}^{2}d\bar{R}^{2}\right)
- \bar{\alpha}^{2}\bar{R}^{2m^{2}}G_{-}^{2}\left(dx^{2}\right)^{2} -
\frac{\bar{R}^{2}G_{-}^{2}}{\bar{C}^{2}}\left(dx^{3}\right)^{2}, \;
(c_{1}c_{2} \not= 0), 
\eq 
where
\bqn
\lb{2.19cc}
\bar{T} &=& \frac{\beta^{m^{2} + 1}}{\gamma^{2}}T,\;\;\; 
\bar{R} = \beta R,\;\;\; 
G_{-} = \bar{R}^{m} - \bar{R}^{-m},\nb\\
\bar{\alpha} &\equiv& \frac{\alpha \gamma}{\beta^{m^{2}}},\;\;\;
\bar{\gamma} \equiv \frac{\gamma}{\beta^{m^{2} + 1}},\;\;\;
\bar{C} \equiv \frac{\beta C}{\gamma}.
\eqn
In the new coordinates $\{\bar{x}^{\mu}\} = \{\bar{T},\; \bar{R},\;
x^{2},\; x^{3}\}$,  the electromagnetic field is given by, 
\bq 
\lb{2.19dd}
\bar{F}^{\mu\nu} = \pm \frac{2m}{\bar{\gamma}^{3}
\bar{R}^{2m^{2}+1}G^{2}_{-}}\left(\delta^{\mu}_{0} \delta^{\nu}_{1}
  - \delta^{\mu}_{1} \delta^{\nu}_{0}\right).
\eq
Therefore, similar to Case B, in this case {\em there are only three physically
independent free parameters}, too. Yet, the solutions also admit the
symmetry $m \leftrightarrow - m$. Thus, in the following we shall consider
only the case where $m \ge 0$ and study these solutions in the $\bar{x}^{\mu}$
coordinates.  Without causing any confusions, we shall drop
all the bars from the above constants and coordinates. Then, it can be
shown that the corresponding quantities $F$ and $I$ are given by   
 \bqn  
\lb{2.20aa} 
F&\equiv&
F^{\alpha\beta}F_{\alpha\beta} = - 
\frac{8m^{2}}{\gamma^{2}R^{2(m^{2}+1)}G_{-}^{4}},\nb\\ 
I &\equiv& R^{\alpha\beta\gamma\sigma} R_{\alpha\beta\gamma\sigma}  =
\frac{16m^{2}}{\gamma^{4}R^{4(m^{2}+1)}G_{-}^{8}}\times\nb\\ & &
\left\{(m+1)^{2}\left[m(m+1) + 1\right] R^{4m} + (m-1)^{2}\left[m(m-1)
+ 1\right] R^{-4m}\right.\nb\\ & & 
\left. + 6m(m+1)^{2}  R^{2m} - 6m(m-1)^{2} R^{-2m} -
2(m^{4} - 12m^{2} + 1)\right\}, 
\eqn 
from which we find that
\bq
\lb{2.21aa}
F = \cases{-\infty, & $m \not=  1$\cr
{- 8\gamma^{-2}}, & $ m =  1$\cr}\;\;,\;\;\;\;\;
I = \cases{\infty, & $m \not=  1$\cr
-{320\gamma^{-4}}, & $ m =  1$ \cr}\;\;,
\eq
as $R \rightarrow 0^{+}$. From the above expressions it can be also shown that
$F$ and $I$ all go to zero as $R \rightarrow + \infty$. Therefore, all these
solutions are asymptotically flat as $R \rightarrow +\infty$ and singular at $
 R = 0$, except for the one with $m = 1$. In addition to the
singularities at $R = 0$, the solutions are also singular at $R =
1$ where $G_{-} = 0$.  Thus, when $m \not=  1$,
the semi-axis $R \ge 0$ is divided into two parts, $ 0 \le R \le 1$ and
$1 \le R < + \infty$,  by the singularities  located,
respectively, at $R = 0$ and $R= 1$. While the physics of the spacetime in
the region  $ 0 \le R \le 1$  is not clear, one can introduce a new
coordinate $R'$ by $R' = R - 1$ in the region $1 \le R < +
\infty$, so the solutions are singular at $R' = 0$ and asymptotically flat as
$R' \rightarrow + \infty$. The spacetime is maximal in this region and the
corresponding Penrose diagram is given by Fig. 1.

When $m = 1$,  the metric (\ref{2.19bb}) takes the form,
\bq
\lb{2.23}
ds^{2} = \gamma^{2}\left[\frac{ R^{2}}{\left(1 - 
R^{2}\right)^{2}}dT^{2} - \left(1 - R^{2}\right)^{2}dR^{2} \right]
 - \left(1- R^{2}\right) \left[\alpha^{2} \left(dx^{2}\right)^{2} + C^{-2}
\left(dx^{3}\right)^{2}\right],\;\; (m = 1), 
\eq
while the electromagnetic field is given by
\bq 
\lb{2.23aaa}
{F}^{\mu\nu} = \pm \frac{2}{{\gamma}^{3}
R(1-R^{2})^{2}}\left(\delta^{\mu}_{0} \delta^{\nu}_{1}   - \delta^{\mu}_{1}
\delta^{\nu}_{0}\right). 
\eq

Introducing the new coordinates $t$ and $r$ via the relations
\bq
\lb{2.24}
t = \frac{1}{2}T,\;\;\; 
r = \frac{1}{2}\left(1 - R^{2}\right),
\eq
we find that the metric takes the form
\bq
\lb{2.23a}
ds^{2} = 2\gamma^{2}\left[f(r)dt^{2} - f^{-1}(r)dr^{2}\right]
 - 4r^{2}\left[\alpha^{2} \left(dx^{2}\right)^{2} + C^{-2}
\left(dx^{3}\right)^{2}\right], \; (m = 1),
\eq
where  
\bq 
\lb{2.25}
f(r) \equiv \frac{r_{0} - r}{r^{2}},
\eq
with $r_{0} \equiv 1/2$. From Eq.(\ref{2.24}) we can see that the
spacetime singularity at $R = 1$ is mapped to $r = 0$, and the
hypersurface $R = 0$ is mapped to $r = r_{0}$. The region
$ 1 \le R < + \infty$ is mapped into the region $r \le 0$. In this region
the singularity at $r = 0$ is naked and timelike, and the corresponding
Penrose diagram is given by Fig. 1. The region $ 0 \le R \le 1 $ is
mapped into $ 0 \le r \le r_{0}$, which will be referred as to Region $I$. The
metric is singular at $r_{0}$. As showed above, this singularity is not a
curvature one, and we need to extend the spacetime beyond it. Since the part 
$$
4r^{2}\left[\alpha^{2} \left(dx^{2}\right)^{2} + C^{-2}
\left(dx^{3}\right)^{2}\right],
$$ 
is regular across $r = r_{0}$,  we need to consider the extension only for
the part
\bq
\lb{2.23b}
d\sigma^{2} =  f(r)dt^{2} - f^{-1}(r)dr^{2}, 
\eq
which is  similar to the Schwarzschild solution with spherical symmetry 
\cite{HE73}. Following the same procedure for the extension of the
Schwarzschild solution, we find that the corresponding Penrose diagram now is
given by Fig. 2.  In this diagram there are three extended regions, $I',
\; II$ and $II'$, where Region $I'$ is symmetric with Region $I$,
while Region $II'$ is symmetric with Region $II$, where in Region
$II$ we have $ r_{0} < r < + \infty$. The spacetime is asymptotically flat
 as $r \rightarrow + \infty$. The nature of the hypersurfaces
$r = r_{0}$, however, is different from   $r = 2m$ in the
Schwarzschild case, representing now   Cauchy
horizons.

In addition to the above solutions, it can be shown that the solution
\bq
\lb{eqa}
f(R) = R^{-2},\;\;\; g(R) = \alpha^{-2}h(R) = R^{2}e^{2\delta R}, 
\;\;\; l(R) = \frac{R^{2}}{C^{2}},
\eq
also satisfies the Einstein-Maxwell equations with the electromagnetic field
being given by
\bq
\lb{eqb}
F^{\mu\nu} = \pm \frac{e^{-2\delta R}}{R^{2}}
\left(\delta^{\mu}_{0}\delta^{\nu}_{1} -
\delta^{\mu}_{1}\delta^{\nu}_{0}\right),
\eq
where $\delta$ is a constant. The corresponding physical quantities $F$ and $I$
are given by 
\bqn
\lb{eqc}
F&\equiv& F^{\alpha\beta}F_{\alpha\beta} = - \frac{2}{R^{4}e^{2\delta R}},
\nb\\ 
I &\equiv& R^{\alpha\beta\gamma\sigma} R_{\alpha\beta\gamma\sigma}  
=  \frac{8}{R^{8}e^{4\delta R}}\left(2\delta^{2}R^{2} + 6\delta R + 7\right),
\eqn 
from which we can see that the spacetime is singular at $R = 0$, and
asymptotically flat as $R \rightarrow \infty$, provided that
\bq
\lb{eqd}
\delta \ge 0,
\eq
a condition that will be imposed in the rest of the paper.

\section{Sources of the LC Solutions  Coupled with
electromagnetic Fields}

As showed in the last section, all the solutions are singular at the axis,
except for the cases $m  = \pm 1$. These singularities are usually considered
as representing sources. However, these sources to be physically acceptable
have to satisfy certain conditions, such as,  the weak, dominant, and$/$or
strong energy conditions \cite{HE73}. Safko and Witten studied several models
of the sources, and found that, when $\sigma \approx 0$, the sources satisfy
some desired physical conditions \cite{SW72}. In this section, we shall
consider shell-like sources.

 Assume that the shell, located on the hypersurface $\Sigma$,
divides the whole spacetime into two regions, $V^{\pm}$. Let  $V^{+}$ denote
the region outside the shell,  and $V^{-}$ denote the region inside the shell.
In $V^{+}$, the  metric takes the form  
\bq 
\label{3.1} 
ds^2_{+} = f^{+}(R){dT}^ 2 - g^{+}(R){dR}^{2} -
h^{+}(R)\left(dx^{2}\right)^{2} - l^{+}(R)\left(dx^{3}\right)^{2},\; 
(R \ge R_{0}),
 \eq
where $\{x^{+\mu}\} = \{T, \; R, \; x^{2}, \; x^{3}\}$, and $R = R_{0} =
Const.$ is the location of the shell in the coordinates $x^{+ \mu}$.  In
$V^{-}$, the  metric takes the form   
\bq
\label{3.2}
ds^2_{-} = f^{-}(r){dt}^ 2 - g^{-}(r){dr}^{2} -
h^{-}(r)\left(dx^{2}\right)^{2} - l^{-}(r)\left(dx^{3}\right)^{2},\; (r \le
r_{0}),
 \eq
where $\{x^{-\mu}\} = \{t, \; r, \; x^{2}, \; x^{3}\}$, and the hypersurface
$r = r_{0}= Const.$ is the location of the shell in the coordinates
$x^{-\mu}$.  On the shell, the intrinsic coordinates will be chosen as 
$\{\xi^{a}\} = \{\tau, \; x^{2}, \; x^{3}\},\; (a = 1,2,3)$, where $\tau$
denotes the proper time of the shell. In terms of $\xi^{a}$, the metric on the
shell takes the form
\bq
\label{3.3}
\left.ds^2\right|_{\Sigma}=  \gamma_{ab}d\xi^{a}d\xi^{b} = {d\tau} ^{2}   -
h\left(dx^{2}\right)^{2} - l\left(dx^{3}\right)^{2},
\eq
where $\gamma_{ab}$ denotes the induced metric on the hypersurface.
The first junction condition requires that the metrics in both sides of the
shell reduce to the same metric (\ref{3.3}), that is, 
\bqn
\label{3.4}
\left[f^{+}(R_{0})\right]^{1/2}dT &=& \left[f^{-}(r_{0})\right]^{1/2}dt =
d\tau,\nb\\ 
h^{+}(R_{0}) &=& h^{-}(r_{0}) = h,\;\;\;
l^{+}(R_{0}) = l^{-}(r_{0}) = l.
\eqn
Note that in writing the above expressions, we have chosen, without loss of
generality, $dT,\; dt$ and $d\tau$ to have the same sign.
The normal vector to the hypersurface $\Sigma$ is given in $V^{+}$ and
$V^{-}$, respectively,  by  
\bq
\lb{3.5}
n^{+}_{\mu} = \left[g^{+}(R_{0})\right]^{1/2}\delta^{R}_{\mu},\;\;
n^{-}_{\mu} = \left[g^{-}(r_{0})\right]^{1/2}\delta^{r}_{\mu}.
\eq
On the hypersurface $\Sigma$, let us introduce the vectors, $e_{(a)}^{\pm\mu}$,
defined by $e_{(a)}^{\pm\mu} \equiv \partial x^{\pm \mu}/\partial \xi^{a}$, we
find that
\bqn
\lb{3.6}
e^{+\mu}_{(1)} = {\left[f^{+}(R_{0})\right]^{-1/2}}
\delta^{\mu}_{T},\;\;\;  e^{+\mu}_{(2)} =  \delta^{\mu}_{2},\;\;\; 
e^{+\mu}_{(3)} =  \delta^{\mu}_{3},\nb\\
e^{-\mu}_{(1)} = {\left[f^{-}(r_{0})\right]^{-1/2}}
\delta^{\mu}_{t},\;\;\;  e^{-\mu}_{(2)} =  \delta^{\mu}_{2},\;\;\; 
e^{-\mu}_{(3)} =  \delta^{\mu}_{3}.
\eqn
Then, the extrinsic curvatures $K^{\pm}_{ab}$, defined by \footnote{Note that
in this paper the definition for the extrinsic curvature tensor is the
same as that given in \cite{PW00} but different from Israel's by a 
sign \cite{Israel66}.} 
\bq 
\lb{3.7}
K^{\pm}_{ab} = - e^{\pm\alpha}_{(a)}e^{\pm\beta}_{(b)}\left\{
\frac{\partial^{2}n^{\pm}_{\alpha}}{\partial \xi^{a} \partial
\xi^{b}} - \Gamma^{\pm\lambda}_{\alpha\beta}n^{\pm}_{\lambda}\right\},
\eq
have the following non-vanishing components,
\bqn
\lb{3.8}
K^{+}_{11} &=&  \frac{f^{+}_{,R}}{2f^{+}(g^{+})^{1/2}},\;\;\;
K^{+}_{22} = - \frac{h^{+}_{,R}}{2(g^{+})^{1/2}},\;\;\;
K^{+}_{33} = - \frac{l^{+}_{,R}}{2(g^{+})^{1/2}},\nb\\
K^{-}_{11} &=&  \frac{f^{-}_{,r}}{2f^{-}(g^{-})^{1/2}},\;\;\;
K^{-}_{22} = - \frac{h^{-}_{,r}}{2(g^{-})^{1/2}},\;\;\;
K^{-}_{33} = - \frac{l^{-}_{,r}}{2(g^{-})^{1/2}}.
\eqn
In terms of $K^{\pm}_{ab}$, the surface energy-momentum tensor, $\tau_{ab}$,
is given by \cite{Israel66},
\bq
\lb{3.9}
\tau_{ab} = \frac{1}{\kappa} \left\{\left[K_{ab}\right]
- \gamma_{ab}\left[K\right]\right\},
\eq
where $\left[K_{ab}\right] \equiv K^{+}_{ab} - K^{-}_{ab}$ and
$\left[K\right] \equiv \gamma^{ab}
\left[K_{ab}\right]$. Substituting Eq.(\ref{3.8}) into Eq.(\ref{3.9}) we
find that $\tau_{ab}$ can be written in the form,
\bq
\lb{3.10}
\tau^{ab} = \rho w^{a}w^{b} + p_{2}e_{(2)}^{a}e_{(2)}^{b} 
+ p_{3}e_{(3)}^{a}e_{(3)}^{b},
\eq
where $w^{a} = \delta_{\tau}^{a},\;
e_{(2)}^{a} = h^{-1/2}\delta_{2}^{a},\;
e_{(3)}^{a} = l^{-1/2}\delta_{3}^{a}$, and
\bqn
\lb{3.11}
\rho &=& - \frac{1}{2\kappa}\left\{
\frac{1}{h}\left[\frac{h^{+}_{,R}}{\left(g^{+}\right)^{1/2}} 
- \frac{h^{-}_{,r}}{\left(g^{-}\right)^{1/2}}\right] 
+ \frac{1}{l}\left[\frac{l^{+}_{,R}}{\left(g^{+}\right)^{1/2}} 
- \frac{l^{-}_{,r}}{\left(g^{-}\right)^{1/2}}\right]\right\},\nb\\
p_{2} &=&  \frac{1}{2\kappa}\left\{
\left[\frac{f^{+}_{,R}}{f^{+}\left(g^{+}\right)^{1/2}} 
- \frac{f^{-}_{,r}}{f^{-}\left(g^{-}\right)^{1/2}}\right] 
+ \frac{1}{l}\left[\frac{l^{+}_{,R}}{\left(g^{+}\right)^{1/2}} 
- \frac{l^{-}_{,r}}{\left(g^{-}\right)^{1/2}}\right]\right\},\nb\\
p_{3} &=&  \frac{1}{2\kappa}\left\{
\left[\frac{f^{+}_{,R}}{f^{+}\left(g^{+}\right)^{1/2}} 
- \frac{f^{-}_{,r}}{f^{-}\left(g^{-}\right)^{1/2}}\right] 
+ \frac{1}{h}\left[\frac{h^{+}_{,R}}{\left(g^{+}\right)^{1/2}} 
- \frac{h^{-}_{,r}}{\left(g^{-}\right)^{1/2}}\right]\right\}.
\eqn
Thus, the surface EMT given above can be considered as representing a fluid
with its velocity $w^{a}$, energy density $\rho$ and pressures $p_{2}$
and $p_{3}$ in the two principal directions $e_{(2)}^{a}$ and $e_{(3)}^{a}$,
respectively, provided that the fluid satisfies some energy conditions 
\cite{HE73}.

Once we have the general formulae for the matching of two static cylindrical
regions, let us consider some specific models, where the solutions given in
the last section are taken as valid only in the region $V^{+}$ defined
above. To make sure that the spacetimes indeed possess cylindrical symmetry,
and that the LC vacuum solutions or the LC solutions coupled with an
electromagnetic field are produced by a cylindrically symmetric source,  in the
region $V^{-}$ we shall choose the metric as that of Minkowskian, 
\bq 
\lb{3.12}
ds^{2}_{-} = dt^{2} - dr^{2} - dz^{2} - r^{2}d\varphi^{2}, \; (r \le r_{0}),
\eq 
so that the spacetime and its symmetry inside the shell is well defined and
free of any kind of spacetime singularities on the axis $r = 0$. Obviously,
for such a matching a matter shell in general appears on the hypersurface $r =
r_{0}$. Since inside the shell, the spacetime is flat and free of any kind of
sources, the spacetime outside the shell is produced solely by the
shell. 

Denoting the electromagnetic field outside the cylinder by $F^{+\mu\nu}$, we
can write it in the whole spacetime as
\bq 
\lb{3.12a}
F^{\mu\nu} = F^{+\mu\nu}H(R-R_{0}),  
\eq 
where $H(R-R_{0})$ is the step function defined by
\bq 
\lb{3.12b}
H(R-R_{0}) \equiv \cases{1, & $R \ge R_{0}$,\cr
0, & $R < R_{0}$.\cr}
\eq 
From Eq.(\ref{2.8}) that is now valid only outside
the shell, we find that in the whole spacetime $F^{\mu\nu}$ defined by the
above equations satisfies the Maxwell equation,
\bq 
\lb{3.12c}
F^{\mu\nu}_{\;\;\; ;\nu} = J^{\mu}, 
\eq
where $J^{\mu}$ is given by
\bq 
\lb{3.12d}
J^{\mu} \equiv F^{+\mu R}\delta(R - R_{0}), 
\eq
with $\delta(R - R_{0})$ being the Dirac delta function. To further study the
problem, let us consider the four cases defined by Eq.(\ref{2.12}) separately.

\subsection{$B^{0} =  B^{2} = B^{3} = 0$}

In this case,  the spacetime outside the shell is described by the LC
solutions, which are given by Eq.(\ref{2.13}) or (\ref{2.13b}). Without loss of 
generality, in the following we shall consider only the metric of Eq.(\ref{2.13}),
which is valid for any $\sigma$. Since the spacetime outside the shell is
vacuum, $F^{+\mu\nu} = 0$, from Eqs.(\ref{3.12a})-(\ref{3.12d}) we can see
that the shell is free of electromagnetic charge and current. 

Because of the ambiguity of specifying the angular 
and axial coordinates, let us first consider the possible identification $z
= x^{2}$  and $\varphi = x^{3}$. In this case the first junction condition of
Eq.(\ref{3.4}) yields 
\bq
\lb{3.13}
\alpha = R_{0}^{-2\sigma(2\sigma - 1)},\;\;\;\;
C = \frac{R_{0}^{1-2\sigma}}{r_{0}}.
\eq
Then, from Eq.(\ref{3.11}) we find that
\bqn
\lb{3.14}
\rho &=& \frac{1}{\kappa}\left[\frac{1}{r_{0}} 
- \frac{(2\sigma -1)^{2}}{R_{0}^{{\cal{A}}}}\right],\nb\\
p_{z} &=& \frac{1}{\kappa}\left(
\frac{1}{R_{0}^{{\cal{A}}}} - \frac{1}{r_{0}}\right),\;\;\;\;\;
p_{\varphi} =  \frac{4\sigma^{2}}{\kappa 
R_{0}^{{\cal{A}}}}.
\eqn
From the above expressions, it can be shown that the weak and strong energy
conditions will be fulfilled when
\bq
\lb{3.14a}
r_{0} \le \frac{R_{0}^{{\cal{A}}}}{(2\sigma -1)^{2}},\;\;\;\;
0 \le \sigma \le 1,
\eq
while the dominant energy condition requires
\bq
\lb{3.14b}
r_{0} \le \cases{\frac{R_{0}^{{\cal{A}}}}{(2\sigma -1)^{2} + 1}, 
& $ 0 \le \sigma \le 1/3$\cr
\frac{R_{0}^{{\cal{A}}}}{(2\sigma -1)^{2} + 4\sigma^{2}}, 
& $ \sigma > 1/3$\cr},\;\;\;\;\;
0 \le \sigma \le 1.
\eq
Clearly, by properly choosing the constant $r_{0}$
the three energy conditions, weak, dominant and strong, can be all satisfied, 
provided that
\bq
\lb{3.15}
0 \le \sigma \le 1.
\eq
That is, the solutions of Eq.(\ref{2.13}) with $z = x^{2}$ and $\varphi =
x^{3}$ can be produced by physically reasonable sources for $0 \le \sigma \le
1$. This  is consistent with the conclusions obtained in \cite{WSS97}. 

When $\sigma$ is larger, the two coordinates $x^{2}$ and $x^{3}$  change 
the roles, as we pointed in the last section. In the following, we shall show that 
this is indeed the case. Choosing $z = x^{3}$ and $\varphi = x^{2}$ in Eq.(\ref{2.13}) 
we find that the first junction  condition of Eq.(\ref{3.4})
becomes
\bq
\lb{3.16}
\alpha = \frac{r_{0}}{R_{0}^{2\sigma(2\sigma - 1)}},\;\;\;\;
C = R_{0}^{1-2\sigma},
\eq
while Eq.(\ref{3.11}) yields 
\bqn
\lb{3.17}
\rho &=& \frac{1}{\kappa}\left[\frac{1}{r_{0}} 
- \frac{(2\sigma -1)^{2}}{R_{0}^{{\cal{A}}}}\right],\nb\\
p_{z} &=&  \frac{1}{\kappa R_{0}^{{\cal{A}}}},\;\;\;\;\;
p_{\varphi} = \frac{1}{\kappa}\left(
\frac{4\sigma^{2}}{R_{0}^{{\cal{A}}}} - \frac{1}{r_{0}}\right).
\eqn
From these expressions, it can be shown that the weak and strong energy
conditions will be satisfied when
\bq
\lb{3.17a}
r_{0} \le \frac{R_{0}^{{\cal{A}}}}{(2\sigma -1)^{2}},\;\;\;\;
 \sigma \ge \frac{1}{4},
\eq
and that the dominant energy condition will be satisfied when
\bq
\lb{3.17b}
r_{0} \le \cases{\frac{R_{0}^{{\cal{A}}}}{(2\sigma -1)^{2} + 1}, 
& $ 1/4 \le \sigma \le 3/4$\cr
\frac{R_{0}^{{\cal{A}}}}{4\sigma(2\sigma -1) + 1}, 
& $ \sigma > 3/4$\cr},\;\;\;\;\;
 \sigma \ge \frac{1}{4}.
\eq
Thus, by properly choosing the constant $r_{0}$ the three energy conditions 
can be all satisfied, for
\bq
\lb{3.18}
\sigma \ge \frac{1}{4}.
\eq
That is, the solutions of Eq.(\ref{2.13}) with $z = x^{3}$ and $\varphi =
x^{2}$ are physically acceptable in the sense that they can be produced by
cylindrical matter shells that satisfy all the three energy conditions,
provided that $\sigma  \ge 1/4$. 

This confirms our early claim that when $\sigma$ is large, the coordinate $x^{2}$
should be taken as the angular coordinate. From 
Eqs.(\ref{3.15}) and (\ref{3.18}) we can see that there exists a common range
$1/4 \le \sigma \le 1$, in which the angular coordinate can be chosen as either
$x^{2}$ or $x^{3}$. For each of such choices, the solutions can be produced
by cylindrical matter shells that satisfy all the three energy conditions.
However, considering Eq.(\ref{2.2}),  we can see that when $\sigma > 1/2$ the
coordinate $x^{2}$ is more likely to play the role of the angular coordinate,
while when $\sigma < 1/2$ the coordinate $x^{3}$  is more likely. When $\sigma
= 1/2$, the corresponding solution becomes (locally) Minkowskian and the metric
coefficients $g_{22}$ and $g_{33}$ are constant. In \cite{SWS98} it was shown
that it can be considered  as representing the gravitational field 
produced by a massive plane with a uniform matter distribution. In this case, the
ranges of the two coordinates $x^{2}$ and $x^{3}$ were extended to $ - \infty < 
x^{2},\; x^{3} < + \infty$. The above considerations, on the other hand, show that
the same solution can be also considered as representing the gravitational field produced
by a cylindrical shell that satisfies all the energy conditions, but in the latter case
one has to identify the hypersurface $x^{2} = 0 \; (x^{3} = 0)$ with the one
$x^{2} = 2\pi \; (x^{3} = 2\pi)$.  

In any case, the above analysis shows clearly that {\em all the LC solutions with}
$\sigma \ge 0$ {\em are physically acceptable}, in the sense that they can be produced by 
cylindrically symmetric sources that satisfy all the three energy conditions \cite{HE73}.

We would like to note that if the form of the metric Eq.(\ref{2.13b}) of the
LC solutions is used as the exterior of the shell, we shall
obtain the same conclusions,  that is, for the solutions to be produced by a
cylindrical shell that satisfies all the  three energy conditions, we have to
choose  $x^{3}$ to be the angular coordinate  $\varphi$ for $0 \le \sigma \le
1$, and to choose $x^{2}$ to be the angular coordinate for $\sigma \ge 1/4$.

\subsection{$B^{2} \not= 0, \; B^{0} = B^{3} = 0$}

In this case, the metric outside the shell is given by Eqs.(\ref{2.19b})  and
(\ref{2.19c}) with all the bars being dropped. 
Let us first consider the case where $x^{2} = z$ and $x^{3} = \varphi$. Then,
we can see that the first junction conditions Eq.(\ref{3.4}) require
\bq
\lb{3.19}
\alpha = G_{+}(R_{0}),\;\;\;\;
C = \frac{R_{0}G_{+}(R_{0})}{r_{0}},
\eq
while Eq.(\ref{3.11}) gives
\bqn
\lb{3.20}
\rho &=& \frac{1}{\kappa}\left(\frac{1}{r_{0}} 
- \frac{2}{\alpha\gamma R_{0}^{1 + m^{2}}}\right),\;\;\;\;\;
p_{\varphi} =  \frac{m^{2}}{\kappa \alpha\gamma R_{0}^{1+m^{2}}},\nb\\
p_{z} &=& \frac{1}{\kappa}\left[
\frac{2m\left(R_{0}^{m} - R_{0}^{-m}\right)}{\alpha^{2}\gamma R_{0}^{1+m^{2}}}
+ \frac{1+m^{2}}{\alpha\gamma R_{0}^{1+m^{2}}} - \frac{1}{r_{0}}\right].
\eqn
From the above expressions we can show that all the three energy conditions
can be satisfied by properly choosing the two constants $r_{0}$ and $R_{0}$,
provided that
\bq
\lb{3.21}
m < - (\sqrt{2} - 1),\;\;\;\;\; {\mbox{or}}\;\;\;\;\;
m >  \sqrt{2} - 1.
\eq
That is, in this case if we make the identification
$x^{2} = z$ and $x^{3} = \varphi$, all the solutions with $m < - (\sqrt{2} - 1)$ or 
with $m >  \sqrt{2} - 1$ can be produced by cylindrically symmetric shells 
that satisfy all the three energy conditions. 

As shown in the last section, the parameter $m$ is related to $\sigma$ via the
relation,
\bq
\lb{3.22}
m = \pm \frac{2\sigma}{2\sigma - 1},
\eq
where the signs ``$\pm$'' depend on the way how to take the vacuum
limits. However, in any case, in terms of $\sigma$, Eq.(\ref{3.21}) takes the
form,
\bq
\lb{3.21a}
\sigma < - \frac{1}{\sqrt{8}},  \;\;\;\;\; {\mbox{or}}\;\;\;\;\;
\sigma >   \frac{1}{\sqrt{8}}.
\eq
Comparing this result with the corresponding 
one for the LC solutions obtained in the last subsection, we can see that 
the coupling of the electromagnetic field with the gravitational  field of
the LC solutions  extends the range $0 \le \sigma \le 1$  to the range 
$\sigma > {1}/{\sqrt{8}}$ or to $\sigma < - {1}/{\sqrt{8}}$. The
extension to the negative values of $\sigma$ is particularly interesting, as
in the vacuum case $\sigma < 0$ corresponds to the situation where the
solutions are produced by negative mass \cite{Bonnor92,WSS97}. 

In this case, the electromagnetic field outside the shell can be considered
as produced by a charge current along it with  zero charge density. 
 This current can be produced, for example,  by moving electrons in the shell,
but the linear charge density of the electrons is equal to the one of ions with
opposite sign, where the ions are at rest. Then, the motion of the electrons
will produce a net  current. This charge current per unit length along the axis
can be calculated by    
\bq  
\lb{3.21a1} 
I_{z} = - \int\int{n_{\lambda}J^{\lambda}
\sqrt{g^{(2)}}dR d\varphi} = \pm \frac{4\pi m}{\gamma C R^{m^{2}}_{0}G_{+}},
\eq
where $n_{\mu} = (-g_{zz})^{1/2}\delta^{z}_{\mu}$ denotes the unit vector
along the axial direction, and $g^{(2)}$ is the determinant of the induced
metric on the 2-surface defined by $t, z =$ Const. The current density
$J^{\mu}$ is defined by Eq.(\ref{3.12d}).

Now let us turn to consider the identification $x^{2} = \varphi$ and $x^{3} =
z$. Then,  the first junction condition Eq.(\ref{3.4}) requires
\bq
\lb{3.23}
\alpha = r_{0}G_{+}(R_{0}),\;\;\;\;
C = R_{0} G_{+}(R_{0}),
\eq
while from Eq.(\ref{3.11}) we find that
\bqn
\lb{3.24}
\rho &=& \frac{1}{\kappa}\left[\frac{1}{r_{0}} 
- \frac{1}{\gamma G_{+}(R_{0}) R_{0}^{1 + m^{2}}}\right],\;\;\;\;\;
p_{\varphi} =  \frac{1}{\kappa}\left[\frac{m^{2}}{\gamma G_{+}(R_{0})
R_{0}^{1+m^{2}}} - \frac{1}{r_{0}}\right],\nb\\ 
p_{z} &=&
\frac{1}{\kappa \gamma G_{+}(R_{0})R^{1+m^{2}}}\left[( 1+ m^{2}) +
\frac{2m}{G_{+}(R_{0})}\left( R_{0}^{m} - 
R_{0}^{-m}\right)\right].  
\eqn 
From these expressions it can be shown that by properly choosing the constant
$r_{0}$, all the three energy conditions can be satisfied, provided that
\bq
\lb{3.25}
m^{2} \ge 1, \;\;\;\;\; {\mbox{or}}\;\;\;\;\;
\sigma \ge   \frac{1}{4}.
\eq
This is the same condition as that given in the corresponding vacuum solutions,
given by Eq.(\ref{3.18}). 

In this case, the shell can be considered as a solenoid, which
produces the electromagnetic field outside the shell. The current per
unit length of the solenoid is given by  
\bq 
\lb{3.21a2}
I_{\varphi} = - \int\int{n_{\lambda}J^{\lambda} \sqrt{g^{(2)}}dR dz}
= \pm \frac{2 m}{\gamma C R^{m^{2}}_{0}G_{+}},
\eq
where now $n_{\mu} = (-g_{\varphi\varphi})^{1/2}\delta^{\varphi}_{\mu}$ denotes
the unit vector along the angular direction, and $g^{(2)}$ is the determinant
of the induced metric on the 2-surface defined by $t, \varphi =$ Const. The
integral is over unit coordinate length along $z$.

\subsection{$B^{3} \not= 0, \; B^{0} = B^{2} = 0 $}

As we noted previously, the case with $B^{3} \not= 0, \; B^{0} = B^{2} = 0$
can be obtained from the case $B^{2} \not= 0, \; B^{0} = B^{3} = 0$ by
exchanging the two spacelike coordinates $z$ and $\varphi$.  Since in the
above,  both possibilities of $(x^{2}, x^{3}) = (z, \varphi)$ and $(x^{2},
x^{3}) = (\varphi, z)$ were considered, the above analysis in fact already
included the case  $B^{3} \not= 0, \; B^{0} = B^{2} = 0$, so in the
following we shall not consider it any more.

\subsection{$B^{0} \not= 0, \; B^{2} = B^{3} = 0$}

In this case, let us first consider the metric outside the shell being given by
Eqs.(\ref{2.19bb}) and (\ref{2.19cc}). As we showed in the
last section, the spacetimes are singular at both $R = 0$ and $ R =
1$. The physics of the spacetimes  in the region $0 \le R \le
1$ is not clear (if there is any), while the spacetimes in the region
$1 < R < + \infty$ are maximal with a naked singularity at $R =
1$. Thus, in this case in order to avoid the presence of spacetime
singularities outside  the shell, we shall assume that $R_{0} > 1$,
where $R_{0}$ denotes the location of the shell in the coordinates $T, \; R,
\; z$, and $ \varphi$. As in the previous cases, now we have two
possibilities of identifying the axial and angular coordinates $z$ and
$\varphi$. Let us first consider the identification  $x^{2} = z$ and $x^{3} =
\varphi$. Then, we find that  the first junction condition Eq.(\ref{3.4})
requires   
\bq 
\lb{3.26}
\alpha = \frac{1}{R_{0}^{m^{2}}G_{-}(R_{0})},\;\;\;\;
C = \frac{R_{0}G_{-}(R_{0})}{r_{0}},
\eq
and that Eq.(\ref{3.11}) gives
\bqn
\lb{3.27}
\rho &=& \frac{1}{\kappa}\left[\frac{1}{r_{0}} 
- \frac{1+m^{2}}{\gamma R_{0}^{1 + m^{2}}G_{-}(R_{0})}
- \frac{2m}{\gamma R_{0}^{1 + m^{2}}G_{-}^{2}(R_{0})}
\left(R_{0}^{m} + R_{0}^{-m}\right)\right],\nb\\
p_{z} &=& \frac{1}{\kappa}\left[
\frac{1}{\gamma R_{0}^{1 + m^{2}}G_{-}(R_{0})} -
\frac{1}{r_{0}}\right],\;\;\;\; 
p_{\varphi} =  \frac{m^{2}}{\kappa \gamma R_{0}^{1 +
m^{2}}G_{-}(R_{0})}. 
\eqn
From these expressions it can be shown that none of the three energy
conditions is satisfied. Thus, unlike the last subcase, all the solutions
in this subcase cannot be produced by physically acceptable thin shells.
Therefore, these models have to be discarded. 

If we choose $x^{2} = \varphi$ and $x^{3} = z$, then we find that  
\bq
\lb{3.29}
\alpha = \frac{r_{0}}{R_{0}^{m^{2}}G_{-}(R_{0})},\;\;\;\;
C = R_{0}G_{-}(R_{0}),
\eq
and that 
\bqn
\lb{3.30}
\rho &=& \frac{1}{\kappa}\left[\frac{1}{r_{0}} 
- \frac{1+m^{2}}{\gamma R_{0}^{1 + m^{2}}G_{-}(R_{0})}
- \frac{2m}{\gamma R_{0}^{1 + m^{2}}G_{-}^{2}(R_{0})}
\left(R_{0}^{m} + R_{0}^{-m}\right)\right],\nb\\
p_{z} &=&  \frac{1}{\kappa \gamma R_{0}^{1 + m^{2}}G_{-}(R_{0})},\;\;\;\;
p_{\varphi} = \frac{1}{\kappa}\left[
\frac{m^{2}}{\gamma R_{0}^{1 + m^{2}}G_{-}(R_{0})} - \frac{1}{r_{0}}\right],
\eqn
from which it can be shown that, similar to the last case, none of the three
energy conditions is satisfied for any choice of the free parameters
involved, and the models have to be discarded, too. 

Now let us turn to consider the solution given by Eq.(\ref{eqa}) as
representing the spacetime outside the shell. We first consider the case where 
$x^{2} = z$ and $x^{3} = \varphi$. Then,  the first junction
condition Eq.(\ref{3.4}) requires   
\bq 
\lb{3.261}
\alpha = \frac{1}{R_{0}e^{\delta R_{0}}},\;\;\;\;
C = \frac{R_{0}}{r_{0}},
\eq
while Eq.(\ref{3.11}) gives
\bqn
\lb{3.271}
\rho &=& \frac{1}{\kappa}\left(\frac{1}{r_{0}} 
- \frac{2 + \delta R_{0}}{R_{0}^{2}e^{\delta R_{0}}}\right),\nb\\
p_{z} &=& - \frac{1}{\kappa r_{0}},\;\;\;\; 
p_{\varphi} =  \frac{\delta}{\kappa   R_{0}} e^{-\delta R_{0}}. 
\eqn
From these expressions it can be shown that none of the three energy
conditions is satisfied. Therefore, these models, as physically acceptable
ones, have to be also discarded. 

It can be shown that, for the identification $x^{2} = z$ and $x^{3} =
\varphi$, the resultant models are also not physically acceptable, as they all
violate the three energy conditions. As a matter of fact, in the present case 
the first junction condition Eq.(\ref{3.4}) requires    
\bq 
\lb{3.262} \alpha = \frac{r_{0}}{R_{0}e^{\delta R_{0}}},\;\;\;\;
C = {R_{0}},
\eq
while Eq.(\ref{3.11}) gives
\bqn
\lb{3.272}
\rho &=& \frac{1}{\kappa}\left(\frac{1}{r_{0}} 
- \frac{2 + \delta R_{0}}{R_{0}^{2}e^{\delta R_{0}}}\right), \;\;\;\; 
p_{z} = 0, \nb\\
p_{\varphi} &=&  \frac{1}{\kappa } 
\left(\frac{ \delta }{R_{0}e^{\delta R_{0}}} - \frac{1}{r_{0}}\right).
\eqn

Therefore, due to the presence of the purely electric field, the solutions 
cannot be produced by physically acceptable cylindrical shell-like sources.   

\section{Concluding Remarks}

The Levi-Civita solutions coupled with an electromagnetic field are usually 
classified into three different families. In the first two, 
the electromagnetic fields are purely magnetic and produced by a current along,
 the spacelike coordinate $x^{2}$ or $x^{3}$,  while in the last
family  it is purely electric and produced by a charge distribution along the
spacelike coordinate $x^{2}$.  In this paper, the local and global 
properties of all these solutions have been studied, and in particular
found that all the solutions have a naked singularity at $R = 0$, except for
the ones with $m = \pm 1$. In the latter case, two solutions are distinguishable,
one, given by Eq.(\ref{2.22}), is free of any kind of spacetime singularities,
and the corresponding spacetime is geodesically complete. The other, given by 
Eq.(\ref{2.23}), has a coordinate singularity at $r = r_{0}$. After being
maximally  extended beyond this hypersurface, it has been found that this
hypersurface actually represents Cauchy horizons. It has been also found that
the solutions that represent the purely electric fields are also singular at a
finite radial distance $R = 1$. In this case one can introduce a new
radial coordinate  $R'= R- 1$,  so that in terms of $R'$ these
singularities occur at $R'= 0$. Then, one can consider $R'= 0$ as the new
axis, and the resultant spacetimes are  asymptotically flat as $R' \rightarrow
 + \infty$ and maximal in the region $0 \le R' < +\infty$ with a naked
singularity on the axis. 
  
The limits of these solutions to vacuum case have been also studied, and found that
such limits are not unique. As a matter of fact, at least there exist two
different ways to take such limits. For each limit one  gets the
Levi-Civita vacuum solutions with different ranges of values for the  parameter
$\sigma$. From such limiting processes,  we have found that when $\sigma
\rightarrow +\infty$,  the metric becomes Minkowskian with $x^{2}$ as the
angular coordinate \cite{Ph96}, while when  $\sigma \rightarrow 0^{+}$, the
metric becomes Minkowskian, too, but now with $x^{3}$ being the angular
coordinate. This observation leads us to believe that at certain value of
$\sigma$, the two spacelike coordinates $x^{2}$ and $x^{3}$ change their
roles. By constructing cylindrical thin shells, we have been able to confirm
our above expectation, that is,  we have found that, if we make the
identification  $(x^{2}, x^{3}) = (z, \varphi)$, the corresponding Levi-Civita
vacuum solutions can be produced by physically acceptable thin shells only
when $0 \le \sigma \le 1$.  However, if we make the identification $(x^{2},
x^{3}) = (\varphi, z)$, the  corresponding Levi-Civita vacuum solutions can be
produced by physically acceptable  thin shells for $ \sigma \ge 1/4$. 

Cylindrically symmetric thin shells for the Levi-Civita solutions
coupled with electromagnetic fields have been also studied, and found that,
in the case of purely magnetic field, due to the coupling of the
magnetic field with the gravitational field, the range, $0 \le \sigma \le 1$,
of  the corresponding vacuum case, has been extended to   $\sigma > 1/\sqrt{8}$
or to $\sigma < - 1/\sqrt{8}$. The latter extension is  very remarkable, as in
the vacuum case it corresponds to the spacetimes  that are produced by negative
masses \cite{Bonnor92,WSS97}. However, in the case of purely electric field, it
has been found that the solutions resulted from both of the two
identifications,  $(x^{2}, x^{3}) = (z, \varphi)$ and  $(x^{2}, x^{3}) =
(\varphi, z)$, cannot be produced by any physically acceptable cylindrical
thin shells.    Although the sources considered in this paper are the most
general cylindrical thin shells, one may still argue that they are still
not general enough, and therefore, it would be very interesting to look for
other kinds of non-shell-like sources for these solutions.

\section*{Acknowledgements}

We would like to express our gratitude to L. Herrera for valuable
discussions and suggestion. We thank the anonymous referees for useful
suggestions.  The financial assistance from FAPERJ (AW, MFAdaS), CAPES
(AYM) and CNPq (AW, NOS) are gratefully acknowledged.


\newpage

\epsfysize=15cm
 \centerline{\epsfbox{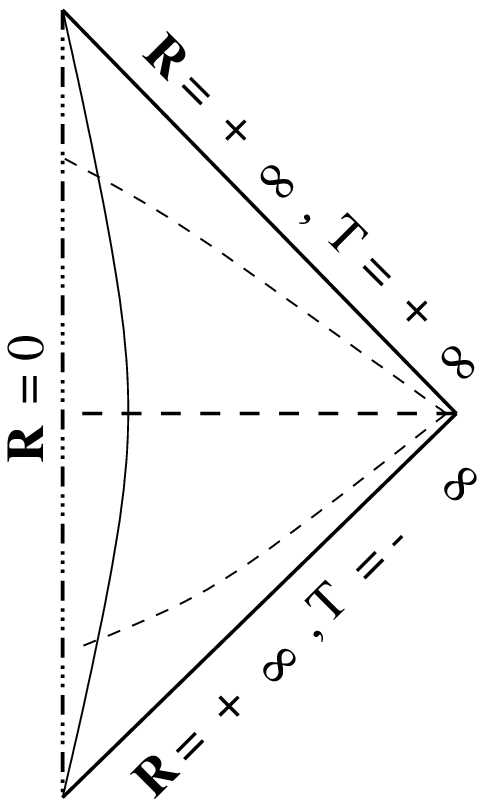}}

\vspace{1.cm}

{\small {Fig. 1 The Penrose diagram. The vertical line $R = 0$ in
general represents a naked spacetime singularity. The dashed lines represent
the hypersurfaces $T = Const.$, and the vertical curved line represents the
hypersurface $R = Const.$}}

\newpage
\epsfysize=15cm
\centerline{\epsfbox{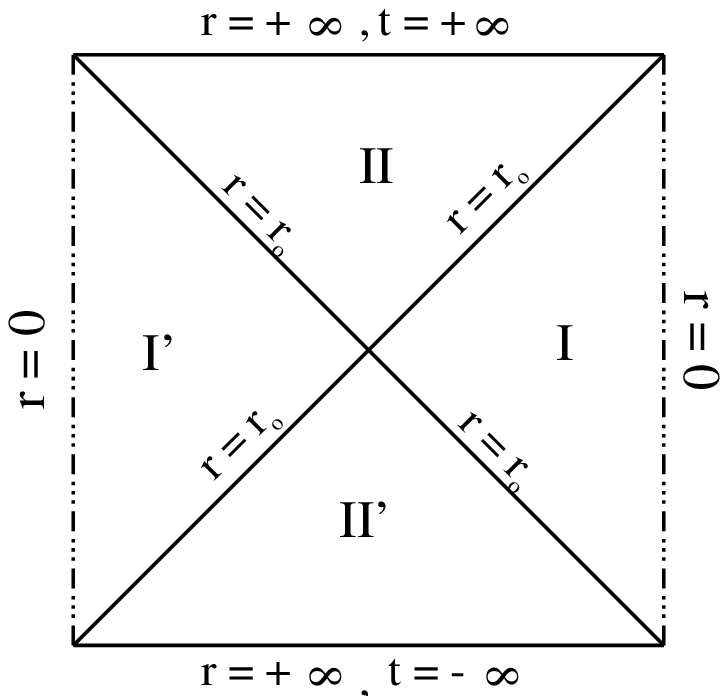}}

\vspace{1.cm}

{\small {Fig. 2 The corresponding Penrose diagram for the solution given by 
Eq.(\ref{2.23a}). The vertical lines $r = 0$ represent the spacetime
singularities that are timelike, and the ones $r = r_{0}$ represent Cauchy
horizons. Region $I$ ($II$) is symmetric to Region $I'$ ($II'$), and Regions
$II$ and $II'$ are asymptotically flat.}}

\end{document}